\documentclass[conference]{IEEEtran}
\IEEEoverridecommandlockouts
\usepackage{amsmath,amssymb,amsfonts}

\usepackage{algorithm}
\usepackage{algorithmic}

\usepackage{graphicx}
\usepackage{textcomp}
\usepackage{xcolor}
\usepackage{keyval}
\usepackage{cite}
\usepackage{graphicx}
\usepackage{caption}

\usepackage{subcaption}
\def\BibTeX{{\rm B\kern-.05em{\sc i\kern-.025em b}\kern-.08em
    T\kern-.1667em\lower.7ex\hbox{E}\kern-.125emX}}
\begin{document}

\title{DQN–Based Joint UAV Trajectory and Association Planning in NTN-Assisted Networks}
\author{
    Afsoon Alidadi Shamsabadi, Cosmas Mwaba, Thomas Nugent, Jie Gao, Pablo Madoery, \\
    Halim Yanikomeroglu, and Subhadeep Pal%
    \thanks{Afsoon Alidadi Shamsabadi, Cosmas Mwaba, Jie Gao, Pablo Madoery, and Halim Yanikomeroglu are with Carleton University, Ottawa, Canada.  Thomas Nugent and Subhadeep Pal are with Honeywell, Ottawa, Canada. Corresponding author Email: \{afsoonalidadishamsa@sce.carleton.ca.\}}
    }
\maketitle

\begin{abstract}
Advanced Air Mobility (AAM) has emerged as a key pillar of next-generation transportation systems, encompassing a wide range of uncrewed aerial vehicle (UAV) applications. To enable AAM, maintaining reliable and efficient communication links between UAVs and control centers is essential. At the same time, the highly dynamic nature of wireless networks, combined with the limited on-board energy of UAVs, makes efficient trajectory planning and network association crucial. Existing terrestrial networks often fail to provide ubiquitous coverage due to frequent handovers and coverage gaps. To address these challenges, geostationary Earth orbit (GEO) satellites offer a promising complementary solution for extending UAV connectivity beyond terrestrial boundaries. This work proposes an integrated GEO–terrestrial network architecture to ensure seamless UAV connectivity. Leveraging artificial intelligence (AI), a deep Q-network (DQN)–based algorithm is developed for joint UAV trajectory and association planning (JUTAP), aiming to minimize energy consumption, handover frequency, and disconnectivity. Simulation results validate the effectiveness of the proposed algorithm within the integrated GEO–terrestrial framework.
\end{abstract}

\section{Introduction}
Advanced air mobility (AAM) represents an emerging paradigm in next-generation air transportation systems, focusing on the safe, efficient, and sustainable movement of people, goods, and services through low-altitude airspace. From passenger air taxis connecting urban centers to uncrewed aerial vehicles (UAVs) conducting cargo delivery, surveillance, and emergency response missions, AAM complements traditional ground-based transportation networks that continue to face growing challenges such as congestion and limited infrastructure capacity~\cite{cel_uav_aam}.
The reliable operation of UAVs to facilitate AAM depends on stable communication links between them and their control centers for command and control signal transmission especially in beyond visual line-of-sight (LoS) operations. To this end, the UAVs can get associated to terrestrial macro base stations (MBSs) which offer multi-point connections for reliability, fast data transfer and allow for long operational distances~\cite{cell_uav}. However, MBSs typically have their antennas down-tilted to provide coverage to ground users, resulting in a weak received signal power~\cite{mob_uav,NesrinArxiv}. Additionally, deploying MBSs in rural and remote areas may be impractical or expensive, leading to coverage gaps and thus limiting UAV operation in such areas. In such situations, non-terrestrial networks (NTN), including the space satellites, can complement terrestrial MBSs in supporting UAV \mbox{connectivity~\cite{ai_sat_uav}}. 

Satellites provide wider geographic coverage over rural and remote regions with few or no MBSs, and supplement coverage and capacity in densely populated urban areas with congested MBSs. By connecting to satellites, UAVs benefit from favorable LoS conditions and enhanced reliability, ensuring consistent communication even in areas beyond terrestrial network coverage. However, UAV operation requires not only a reliable communication link but also demands energy-aware trajectory design, given the UAV’s limited on-board energy~\cite{energy_aware}. Moreover, since UAVs may need to switch between terrestrial and satellite links or among MBSs, minimizing the number of handovers is essential to prevent the ping-pong effect and associated link degradation~\cite{Mozaffari}. Therefore, efficient trajectory and association planning algorithms should aim to minimize handovers along the UAV’s trajectory~\cite{nesrin}.

Artificial intelligence (AI)–driven algorithms, particularly reinforcement learning, have been extensively utilized to address UAV trajectory planning optimization problem. In this context, the authors in~\cite{coverage} and~\cite{transfer} employ deep reinforcement learning (DRL) to enhance UAV communication and control, focusing on objectives such as maximizing energy efficiency and enabling rapid policy adaptation in dynamic environments. 
% Other studies, including~\cite{cargo_aav, balance, jp_connect}, integrate traditional optimization techniques, such as simulated annealing, ant-colony optimization, and genetic algorithms, with DRL to jointly optimize pickup-delivery sequencing and trajectory design, thereby minimizing outage duration and energy consumption. Furthermore, t
The authors in~\cite{nesrin} investigate cargo-delivery applications and apply a Q-learning algorithm to jointly design UAV trajectory and cell association in a standalone terrestrial network. In~\cite{sat_multi_uav1}, multi-agent DRL was employed for trajectory planning of multiple UAVs to maximize ground-user throughput, while the authors in~\cite{sat_uav2} proposed a hybrid graph-neural-network–Q-learning framework to optimize UAV trajectories and satellite link selection for improved user coverage.

This paper investigates an integrated geostationary Earth orbit (GEO) satellite–terrestrial network in which an UAV flies between two designated points. At any given time, the UAV can associate with either the GEO satellite or one of the terrestrial MBSs. Accordingly, a joint UAV trajectory and association planning (JUTAP) problem is formulated to minimize the UAV’s energy consumption, handover, and disconnectivity. To address this problem, a deep Q-network (DQN)–based algorithm is proposed, where the UAV learns optimal trajectory and association policies through interactions with a known environment model. The performance of the proposed DQN framework is then evaluated through simulations under near-realistic satellite and terrestrial channel conditions.

The remainder of this paper is structured as follows. Section \ref{SystemModel} presents the system model and formulates the problem. Section \ref{ProposedAlgorithm} presents the proposed DQN algorithm, and Section \ref{Results} validates the performance of the proposed algorithm. Finally, Section \ref{Conclusion} concludes our paper.

\section{System Model and Problem Formulation}\label{SystemModel}
\begin{figure}[!t]  % !t tries to place it at the top of the page
    \centering
    % Set height; width adjusts automatically to keep aspect ratio
    \includegraphics[width=\linewidth]{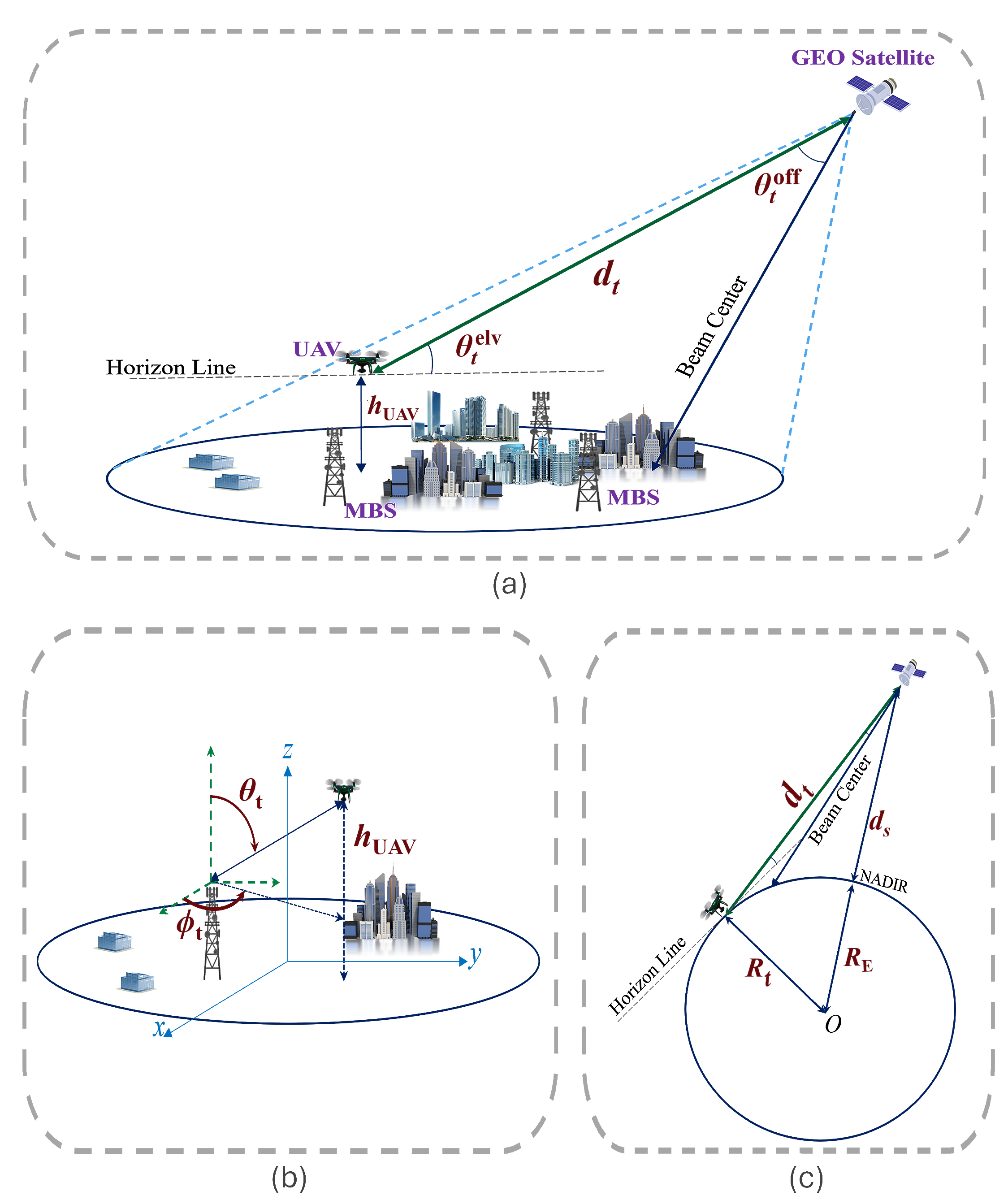}
    \caption{\small System model.}
    \label{fig:sys_mod}
\end{figure}

We consider a geographical region of size $A \times A$, covered by $N_B$ tri-sector multi-antenna MBSs and a single GEO high-throughput satellite that remains stationary relative to the Earth, as illustrated in Fig.~\ref{fig:sys_mod}(a). The UAV is served by either the GEO satellite or one of the MBSs at any given time, depending on the link quality. The GEO satellite is serving the users on the ground and the aerial users using $N$ fixed beams, each covering multiple users in a time division multiplexing manner. The UAV is assumed to be served by one of the GEO beams, utilizing a total bandwidth of $B_G$. The GEO satellite and terrestrial MBSs operate in distinct frequency bands; hence, no inter-tier interference exists between them. Each MBS allocates a total bandwidth of $B_M$ to serve the UAV, while sharing the same time–frequency resources across the terrestrial tier. Consequently, the UAV experiences co-channel interference from neighboring MBSs.

The UAV’s motion is defined within an airspace represented by a three-dimensional (3D) Cartesian coordinate system $(x, y, z)$ for local navigation and an Earth-centered, Earth-fixed (ECEF) geocentric coordinate system for global referencing. The UAV begins its mission from a randomly selected $\text{START}$ location $(x_s, y_s, 0)$ within the coverage area and flies toward a fixed $\text{END}$ position $(x_e, y_e, 0)$. At the $\text{START}$, the UAV vertically ascends to its operational altitude $h_{\text{UAV}}$, constrained to be less than $300$ meters as defined by the Third Generation Partnership Project (3GPP)~\cite{3gpp_tr36777}. This altitude is maintained throughout the flight, and upon reaching the destination, the UAV descends vertically to the ground. 
The $b$-th MBS, $\forall b \in \mathcal{B}=\{1,\dots, N_B\}$, is located at coordinates $(x^{\text{MBS}}_{b}, y^{\text{MBS}}_{b}, z^{\text{MBS}}_{b})$, where $z^{\text{MBS}}_{b}$ denotes the antenna height of the MBS.
We assume that the geographical region is discretized in two dimensions, uniformly divided into $N_G \times N_G$ grid cells, where the $j$-th grid center point $(x_j,y_j,h_{\text{UAV}})$ represents a possible UAV position. We assume that at each time step $t$, the UAV moves from one grid center to another grid center. The discrete trajectory of the UAV is denoted by $\mathcal{T} = \{q_1, \ldots, q_T\}$, where $q_t,~\forall t=1,\dots,T,$ represents the grid index of the UAV position at the $t$-th time step, and $T = |\mathcal{T}|$, corresponds to the total number of discrete time steps in the UAV path. 

We represent the channel gain between the GEO satellite and UAV and the MBS $b,~\forall b \in \mathcal{B}$, and the UAV, at time step $t$, as $h^G_t$ and $h^b_t$. These channel gains are modeled as explained below.
We assume that each sector of MBS $b$ is equipped with uniform planar array comprising $N_A=N^V_A \times N^H_A$ antenna elements, where $N^V_A$ and $N^H_A$ denote the number of vertical and horizontal antenna elements. The array is down-tilted by an angle $\theta_d$; thereby serving the UAV by its side-lobes.
We consider the probabilistic LoS connection between MBSs and the UAV, where the pathloss between the MBS $b$ and UAV, at time $t$, is denoted by $L^b_t$, and is modeled according to~\cite{3gpp_tr36777}. Accordingly, the antenna gain of the MBS $b$ toward UAV at position $(x_t, y_t, h_{\text{UAV}})$, defined as $G^{b}(\theta_t,\phi_t)$, is modeled according to~\cite{ITUR_M2101}, where $\theta_t$ and $\phi_t$ are the elevation and azimuth angles of the UAV with respect to MBS $b$, as depicted in Fig.~\ref{fig:sys_mod}(b). Consequently, the $h^b_t$ will be $\sqrt{{G^{b}(\theta_t,\phi_t)G_r}/{L^b_t}}$, where $G_r$ is the antenna gain at the UAV.

In this system model, the UAV can be reasonably approximated as an elevated user with an almost guaranteed LoS condition to the satellite. To this end, the channel gain between the GEO satellite and UAV, at time step $t$, can be modeled as $h^G_t = \sqrt{{G_r G(\theta^{\text{off}}_t)/L^{G}_t}}$, according to \cite{j_allocation}. In this model, the $G(\theta^{\text{off}}_t)$ is the beam gain of the GEO satellite towards the UAV, positioned at $\theta^{\text{off}}_t$ angle with respect to the GEO's beam center, as shown in Fig.~\ref{fig:sys_mod}(c). The angle $\theta^{\text{off}}_t$ is a function of UAV's location at time step $t$. Additionally, $L^{G}_t$ represents the total propagation loss in the GEO-UAV link, including the free-space path loss, atmospheric loss, scintillation loss, and polarization mismatch loss.
% In the ECEF reference frame, the UAV position is described by its distance from the Earth’s center $R_u$, the distance between the UAV and the satellite $d_u$, and its elevation angle $\theta_{\text{elv}}$ with respect to the satellite. Given that the UAV altitude is significantly smaller than the Earth’s radius ($h_{uav} \ll R_E$), we approximate $r_u \approx R_E$. Furthermore, since we consider a GEO satellite that is fixed relative to the Earth and because the UAV’s horizontal displacement is negligible compared to $R_u$ and $d_u$, the elevation angle $\theta_{\text{elv}})$ can be assumed constant during the UAV's entire flight. We also note that while a GEO satellite is fixed relative to a point on Earth's surface, the UAV's movement across the surface means its elevation angle will change slightly over the flight. However, because the horizontal displacement of the UAV's is very small compared to the orbital distance of the satellite, we can approximate the elevation angle \(\theta _{\text{elv}}\) as constant for simplification purposes.

Considering the MBS-UAV and GEO-UAV channel models, the received signal-to-noise-plus-interference ratio~(SINR) of the UAV from the MBS $b$, at time step $t$, can be expressed as

\begin{equation}
\mathrm{SINR}^b_t = \frac{P^{\text{M}}|h^{\text{b}}_t|^2}
{{\displaystyle\sum_{\substack{j=1,\\ j \neq b}}^{N_B} P^{\text{M}}|h^{\text{j}}_t}|^2 + N_{0} B_{M}},~\forall b \in \mathcal{B}.
\end{equation}
where $P^{\text{M}}$ is the MBS maximum available transmit power and $N_0$ denotes the additive white Gaussian noise (AWGN) power spectral density.
Similarly, the received signal-to-noise ratio (SNR) from the GEO satellite, at time step $t$, can be formulated as
\begin{equation}\label{GEO-SNR}
\mathrm{SNR}^G_t = \frac{p^{\text{G}} \left|h^G_t \right|^2}{ N_0 B_G},
\end{equation}
where $p^{\text{G}}$ is the satellite's maximum available transmit power. Throughout the trajectory, the UAV may experience a coverage hole. The coverage hole, independent of the UAV’s association and trajectory, is defined as the set of grid points where the received SINRs from all MBSs and the SNR from the GEO satellite at the grid center fall below the threshold~$\beta$.

\subsection{UAV Energy Consumption Model}
We consider the energy consumption model for a rotor-wing UAV given in~\cite{rotor_energy}. The UAV propulsion energy required for flying the distance $d$ is defined as $E(\nu, d) = {P_{prop}(\nu)/ d}{\nu}$, where $\nu$ is the UAV's velocity, and $P_{prop}(\nu)$ is the UAV propulsion power given by \cite[Sec.~II.B, Eqn.~12]{rotor_energy}. Consequently, the remaining energy available for the UAV to travel from the START to END grids along trajectory $\mathcal{T}$, with total length $d_\mathcal{T}$, is expressed as $E_A (\nu, d_\mathcal{T}) = E_C - E_S - 2E(\nu, h_\text{UAV})$~\cite[Sec.~II.C, Eqn.~3]{nesrin}, where $E_C$ and $E_S$ represent the initial UAV available energy capacity and the emergency reserve energy, respectively.

\subsection{Formulated JUTAP Problem}\label{ProblemFormulation}
Now, we formulate the JUTAP optimization problem for an UAV traveling from $\text{START}$ to $\text{END}$ on the $N_G \times N_G$ grid. The UAV connectivity along the trajectory $\mathcal{T}$ is characterized by the association sequence $\mathcal{C} = \{c_1,\ldots, c_T\}$, where $c_t$ denotes the association indicator at time step $t$, defined as
\begin{equation}\label{C_t}
\small
    c_{t} =
    \begin{cases}
    b, & \text{if the UAV is associated with MBS $b$,}\\
    N_B+1, & \text{if the UAV is in a coverage hole.}\\
    N_B+2, & \text{if the UAV is associated with the GEO satellite.}
\end{cases}
\end{equation}
    
We assume that at time step $t$, the UAV selects a movement direction $m_t \in \{1, \dots, 8\}$, corresponding to left, right, up, down, up-right, up-left, down-right, and down-left, respectively. The UAV's motion between two consecutive grids $q_t$ and $q_{t+1}$ incurs an energy consumption denoted by $E(v, d_{t,(t+1)})$, where $d_{t,(t+1)}$ is the Euclidean distance between the center of grids $q_t$ and $q_{t+1}$. 
At each transition, a handover cost $\eta_{t,(t+1)}$ is introduced when the UAV changes its serving node, and a disconnectivity indicator $\delta_{t, t+1}$ is imposed when the UAV experiences a coverage hole. 
The weights $w_1$, $w_2$, and $w_3$ are non-negative coefficients representing the relative importance of energy consumption, handover cost, and connectivity reliability, respectively.

Considering the above definitions, the JUTAP problem is formulated as follows:
\begin{IEEEeqnarray}{cl}\label{Problem}
\underset{\mathcal{T},\, \mathcal{C}}{\min} \; &
\sum\limits_{t=1}^{T} 
w_1 E(v, d_{t,(t+1)})
+ w_2 \eta_{t,(t+1)}
+ w_3 \delta_{t,(t+1)}\\
\text{s.t.} \; &
\sum\limits_{t=1}^{T} E(v, d_{t,(t+1)}) \leq E_A(v,h).
\IEEEyessubnumber\label{eq:P1a}
\end{IEEEeqnarray}

The objective of problem \eqref{Problem} is to jointly design the UAV's trajectory $\mathcal{T}$ and association strategy $\mathcal{C}$ such that energy consumption, handover cost, and disconnectivity are jointly minimized while satisfying the UAV's energy constraint.
Problem~\eqref{Problem} is a mixed-integer non-convex optimization problem involving discrete movement direction decisions and association variables. 
Classical Q-learning algorithm can address such sequential decision-making problems; but, it suffers from the curse of dimensionality in large-scale state spaces. 
When the environment consists of a large number of grids and multiple connectivity states, the size of the Q-table becomes extremely large, resulting in slow convergence and poor generalization. 
To overcome these limitations, we adopt a DQN approach, where a neural network is used to approximate the state-action value function, $Q(s,a)$, at each state. The DQN enables efficient policy learning in high-dimensional environments by leveraging experience replay and target network stabilization mechanisms, thus allowing the UAV to learn a trajectory and association policy in a computationally feasible manner.

\section{Proposed DQN Algorithm}\label{ProposedAlgorithm}
Considering the association index $c_t$ defined in \eqref{C_t}, the number of possible associations at time step $t$ increases exponentially with the number of MBSs, resulting in a very large action space for the DQN agent, which significantly complicates the learning process. 
To address this challenge, we simplify the association decision by introducing a binary variable, $\hat{c}_t$, representing the association mode: $\hat{c}_t=0$ indicates terrestrial network association, while $\hat{c}_t = 1$ indicates GEO satellite association. When $\hat{c}_t=0$, the UAV is associated to the MBS offering the maximum received SINR at grid $q_t$. Accordingly, we define the state-action-reward for the DQN algorithm as follows.

\textit{State space:}  
The state at time step $t$ is represented by $s_t = [x_t,\, y_t,\, \hat{c}_t],$ where $(x_t, y_t)$ denote the normalized coordinates of grid $q_t$, and $\hat{c}_t \in \{0,1\}$ is the binary association indicator $\hat{c}_t$. This encoding captures both spatial and association information of the UAV at each time step.

\textit{Action space:}  
At state $s_t$, the agent selects an action $a_t = (m_t, \hat{c}_{t+1})$, where $m_t$ denotes one of the eight possible motion directions, and $\hat{c}_{t+1}$ specifies the binary association index for UAV location at time step $t+1$. The resulting action dimension is thus $|A| = 8 \times 2 = 16$. 

\textit{Reward function:}  
The reward function is designed to minimize the three objective terms for energy consumption, handover cost, and disconnectivity. The immediate reward at time step $t$ is given by
\begin{equation}
r_t = - w_1 \xi_{t,(t+1)} - w_2 \eta_{t,(t+1)} - w_3 \delta_{t,(t+1)},
\label{eq:reward_function}
\end{equation}
where:
\begin{itemize}
    \item $\xi_{t,(t+1)}$ denotes the reward component related to energy consumption, defined as
    \begin{equation}\label{Energy_Reward_Component}
        \xi_{t,(t+1)} =
        \begin{cases}
        -1-d_{\text{norm}}^{(t+1),\text{END}}, & m_t \leq 4, \\
        -1/\sqrt{2}-d_{\text{norm}}^{(t+1),\text{END}}, & \text{Otherwise,}
        \end{cases}
    \end{equation}
    where $d_{\text{norm}}^{(t+1),\text{END}}$ denotes the normalized distance between the grid center at time step $t$ and the END grid.
    \item $\eta_{t,(t+1)}$ represents the handover cost, designed to discourage unnecessary handovers, avoid coverage holes, and prioritize terrestrial connectivity over GEO association due to the higher latency and resource cost of satellite links. The handover cost, $\eta_{t,(t+1)}$, is defined as
    \begin{equation}\label{HandoverCost}
    \small
    \eta_{t,(t+1)} =
        \begin{cases}
        1, & \text{MBS-to-MBS or GEO-to-MBS handover},\quad \\[4pt]
        5, & \text{MBS-to-GEO handover}, \\[4pt]
        0.5, & \text{Remain connected to GEO}, \\[4pt]
        -0.5, & \text{Reconnect after a coverage hole.}
        \end{cases}
    \end{equation}

    These values can be adjusted based on network design preferences and operational requirements.
    \item $\delta_{t,(t+1)}$ denotes the disconnectivity indicator with the values, defined as
    \begin{equation}\label{Disconnectivity_Cost}
    \delta_{t,(t+1)} =
    \begin{cases}
    1, & \text{if $q_{t+1}$ is a coverage hole,}\\
    0, & \text{Otherwise.}
    \end{cases}
    \end{equation}
\end{itemize}

The weighting factors $w_1$, $w_2$, and $w_3$ control the trade-off between energy consumption, handover cost, and disconnectivity, respectively.  
\begin{algorithm}[t]
\caption{Proposed DQN-Based Joint UAV Trajectory and Association Planning (JUTAP) Algorithm}
\label{alg:DQN_UAV}
\begin{algorithmic}[1]
\REQUIRE $(w_1, w_2, w_3)$, $N_{\text{ep}}$, $N_{\text{step}}$, $\gamma$, $\lambda$, $\tau$, $L_B$, $L_\text{Batch}$, $(\epsilon_0, \epsilon_{\min}, \lambda)$.
\ENSURE Trained policy network $\pi_\theta$.
\\[4pt]
\STATE Initialize policy network $Q_\theta(s,a)$ and target network $Q_{\theta^-}(s,a)$ with identical weights.
\STATE Initialize replay buffer with capacity $L_B$.
\FOR{each episode $e = 1, \ldots, N_{\text{ep}}$}
    \STATE Randomly initialize UAV starting grid $q_0$ and initial association $\hat{c}_0 = 0$.
    \STATE Initialize total episode reward $R_e = 0$.
    \FOR{each step $t = 1, \ldots, N_{\text{step}}$}
        \STATE Observe current state $s_t = [x_t, y_t, \hat{c}_t]$.
        \STATE Select action $a_t$ using $\epsilon$-greedy policy, execute $a_t$, obtain next state $s_{t+1}$ and reward $r_t$.
        \STATE Store transition $(s_t, a_t, r_t, s_{t+1})$ into buffer.
        \STATE Sample random mini-batch $\{(s_i, a_i, r_i, s_{i+1})\}_{i=1}^{L_\text{Batch}}$ from buffer.
        \STATE Compute target values $y_i = r_i + \gamma \max_{a'} Q_{\theta^-}(s_{i+1}, a')$.
        \STATE Compute loss function $L(\theta)$, and update $\theta \leftarrow \theta - \eta \nabla_\theta L(\theta)$ via gradient descent.
        \STATE Soft update target network.
        \STATE Update $s_t \leftarrow s_{t+1}$, $\hat{c}_t \leftarrow \hat{c}_{t+1}$, and $R_e \leftarrow R_e + r_t$.
        \IF{UAV reaches destination or $t = N_{\text{step}}$}
            \STATE \textbf{break}
        \ENDIF
    \ENDFOR
    \STATE Decay exploration rate: $\epsilon \leftarrow \max(\epsilon_{\min}, \epsilon_0 \times \lambda^e)$.
\ENDFOR
\end{algorithmic}
\end{algorithm}

The proposed DQN algorithm approximates the optimal state-action value function $Q(s_t, a_t; \theta)$ using a feed-forward neural network.  
The agent follows an $\epsilon$-greedy exploration strategy with an exponentially decaying exploration rate $\epsilon_t = \max(\epsilon_{\min}, \epsilon_0 \times \lambda^t)$.  
At each time step $t$, the agent stores transitions $(s_t, a_t, r_t, s_{t+1})$ in a replay buffer of size $L_B$ and updates the network parameters via mini-batch sampling.  
The loss function is defined as $L(\theta) = \mathbb{E}\big[(r_t + \gamma \max_{a'} Q_{\text{target}}(s_{t+1}, a'; \theta^-) - Q(s_t, a_t; \theta))^2\big]$, where $\gamma$ is the discount factor, and $\theta^-$ are the parameters of the slowly updated target network, updated via soft updates $\theta^- \leftarrow \tau \theta + (1-\tau)\theta^-$.
The proposed DQN algorithm is outlined in Algorithm~\ref{alg:DQN_UAV}. The UAV is initialized at a random grid location at the beginning of each episode and navigates until it reaches the destination $\text{END}$ or the maximum number of time steps, $N_\text{step}$, is reached. For each weight combination $(w_1, w_2, w_3)$, the DQN agent learns a distinct trajectory and association policy that minimizes the total cost in \eqref{eq:reward_function}. 

\section{Simulation Results}\label{Results}
In this section, we evaluate the performance of the proposed DQN-based JUTAP algorithm through simulations. We consider a square region of size $3~\text{km} \times 3~\text{km}$, divided into $25 \times 25$ grids and covered by one GEO satellite and $10$ tri-sector terrestrial MBSs, each sector equipped by $8\times 8$ antenna arrays down-tilted by $\theta_d=10^\circ$. 
The MBSs are spatially distributed in the region according to a homogeneous Poisson point process (PPP) with a minimum separation distance of $500~\text{m}$. Each MBS has a randomly selected height of $35$~m or $25$~m for rural macro and urban macro, respectively. We assume that a GEO satellite positioned at $111.1^{\circ}$W (Anik F2), operating over Ottawa ($45.4^{\circ}$N, $75.7^{\circ}$W), serves the UAV through spot beam $19$ centered near $50^{\circ}$N, $70^{\circ}$W. Due to the region’s small size relative to the satellite’s altitude, the $\theta^\text{off}_t$ is nearly constant, resulting in an almost uniform SNR from the GEO.
The UAV flies at an altitude of $150~\text{m}$ with a constant velocity of $30~\text{m/s}$, starting from the $\text{START}$ point and moving toward the $\text{END}$ point.
Other simulation parameters are summarized in Table~\ref{tab:SimulationParameters}.
\begin{table}
\caption{Simulation Parameters.}\label{tab:SimulationParameters}
\centering
\begin{tabular}{|c||c|}
\hline
\textbf{Parameter} & \textbf{Value}\\
\hline
MBS/GEO carrier frequency & $2.545$ GHz / $2.1$ GHz \\
\hline
$P^M,~P^G$ & $43~\text{dBm},~30~\text{dBm}$\\
\hline
GEO maximum beam gain & $51~\text{dBi}$\\
\hline
AWGN power spectral density, ~$N_{o}$ & $-173.9~\text{dBm/Hz}$\\
\hline
SINR/SNR threshold,~$\beta$ & $-1~\text{dB}$ \\
\hline
\# of hidden layers, Hidden layer size & $2,~128$\\
\hline
Output layer activation function& ReLU\\
\hline
$\gamma,~\tau,~\lambda,~\epsilon_\text{min},~\epsilon_0$ & $1,~0.005,~0.999,~0.1,~1$\\
\hline
$N_\text{step}$,~$N_\text{ep}$ & $1000,~10000$\\
\hline
\end{tabular}
\end{table}
\begin{figure}[t]
    \centering
    \begin{minipage}{\linewidth}
        \centering
        \includegraphics[width=0.95\linewidth]{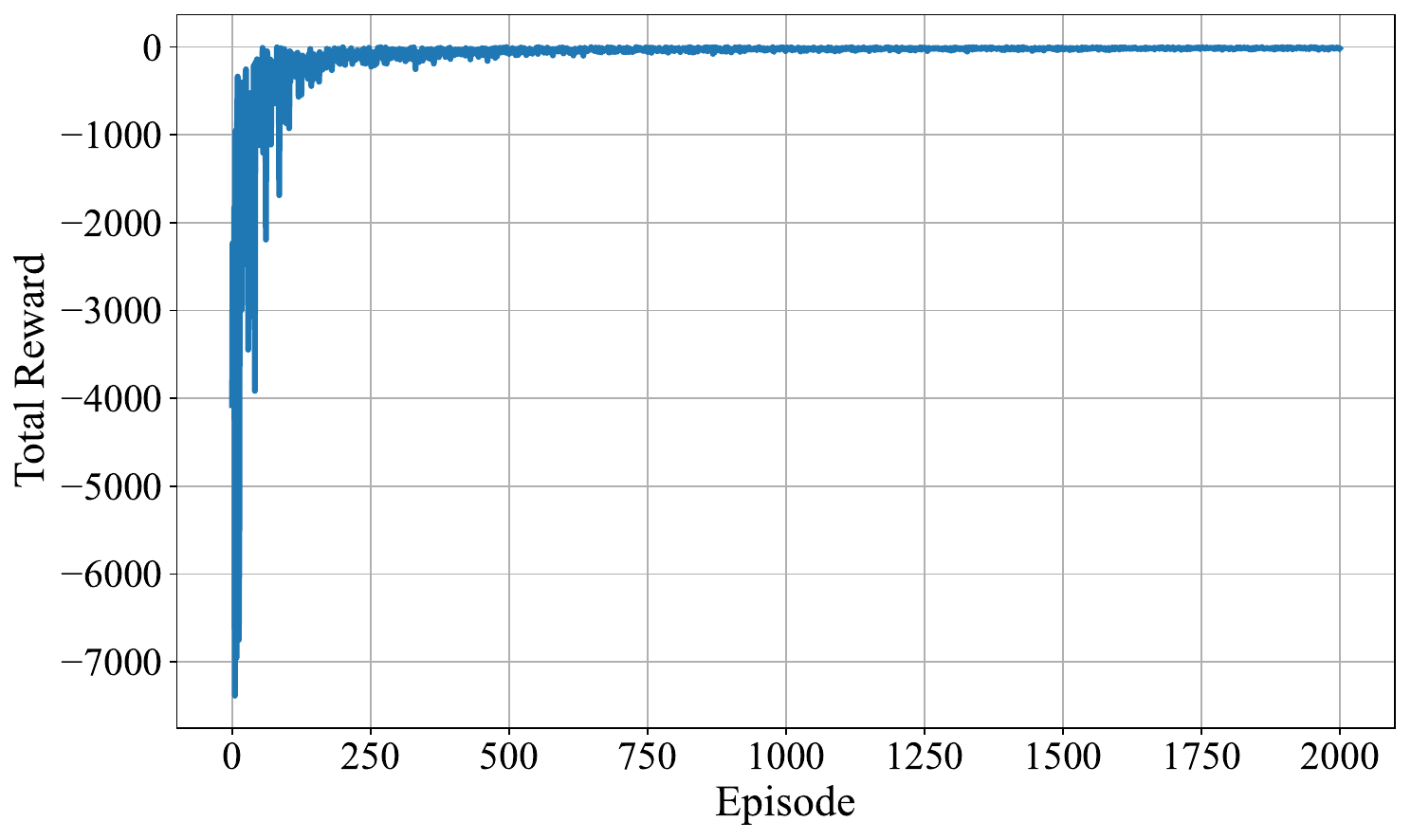}
        \captionof{figure}{\small Total rewards vs episode.}
        \label{fig:Reward}
    \end{minipage}
\end{figure}

Figure.~\ref{fig:Reward} presents the convergence behavior of the proposed DQN algorithm. It can be observed that the total reward gradually increases with training episodes and stabilizes after approximately 400 episodes, demonstrating a stable convergence trend. The early episodes show large reward fluctuations as the agent explores the environment and learns the impact of its actions on the accumulated return. As the learning process continues, the agent achieves a higher reward which is a weighted sum of energy consumption, handover cost, and disconnectivity terms. The steady-state convergence indicates that the proposed DQN model successfully captures the underlying network dynamics and learns the policy for UAV trajectory and association planning.

\begin{figure}[t]
    \centering
    \begin{minipage}{\linewidth}
        \centering
        \includegraphics[width=1.1\linewidth]{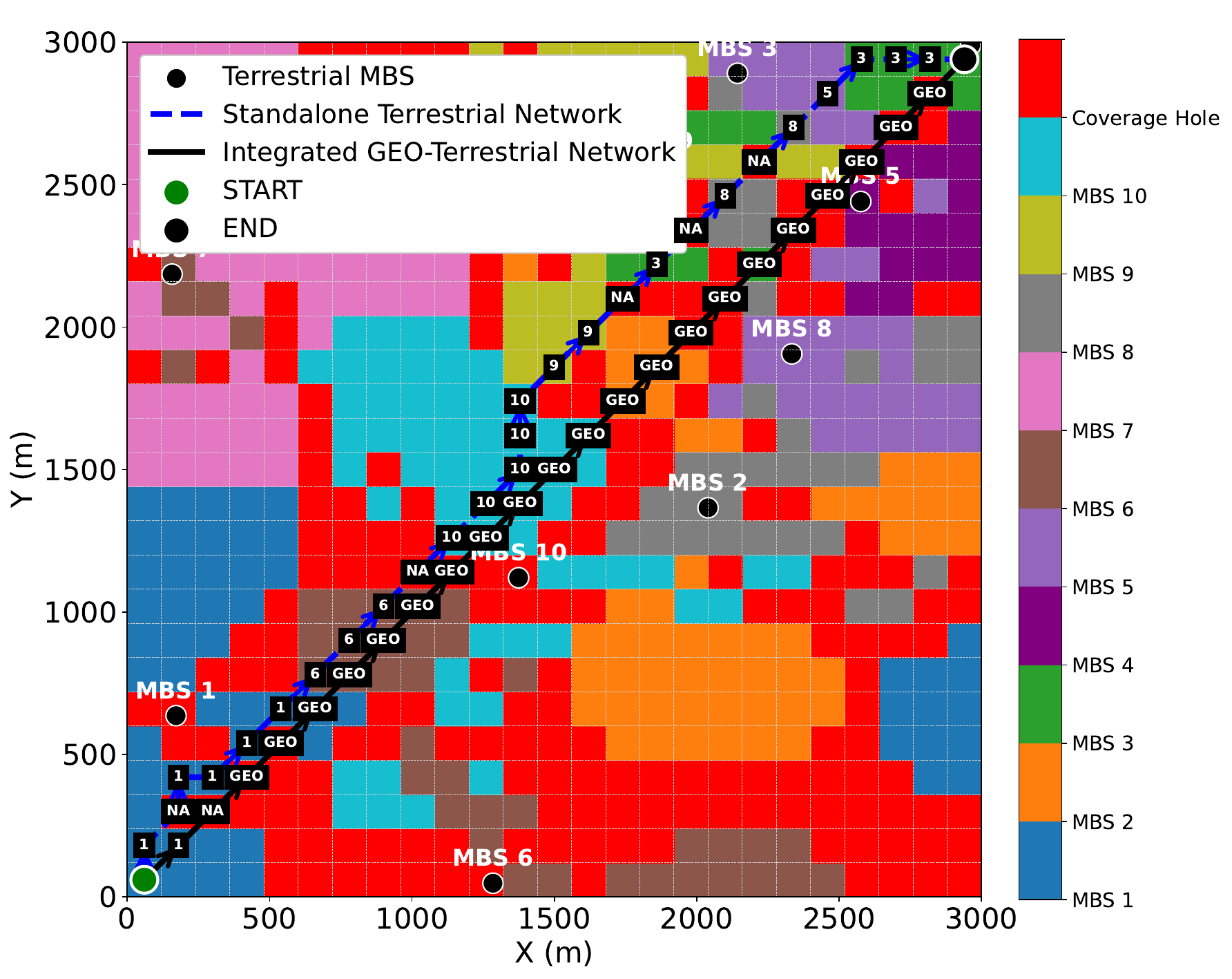}
        \captionof{figure}{Comparison of proposed DQN algorithm in an integrated GEO-terrestrial network and a standalone terrestrial network in a $25 \times 25$ grid ($w_1=0.4,~w_2=0.2,~w_3=0.4$).}
        \label{fig:vHetNetvsTerrestrial}
    \end{minipage}
\end{figure}

Figure.~\ref{fig:vHetNetvsTerrestrial} compares the performance of the proposed DQN algorithm under two scenarios, namely the integrated GEO–terrestrial network and the standalone terrestrial network. In both scenarios, the START and END grids are selected as the bottom-left and top-right grids, respectively. The color assigned to each grid represents the MBS that provides the highest SINR in that location, while red denotes a coverage hole where the SINRs from all MBSs fall below the threshold. Each color in the color bar corresponds to one MBS, with red specifically reserved for coverage holes. Several observations can be made by examining the UAV trajectories in these two cases. First, it can be observed that the integrated GEO–terrestrial network provides the UAV with the opportunity to follow a shorter trajectory by associating with the GEO satellite when beneficial. Second, the UAV can associate with the GEO satellite in coverage holes, thereby maintaining connectivity and avoiding disconnection events. Third, maintaining association with the GEO satellite helps the UAV to avoid frequent handovers within the terrestrial network. According to Fig.~\ref{fig:vHetNetvsTerrestrial}, the GEO-assisted DQN path exhibits a more direct trajectory toward the destination, demonstrating the effectiveness of the integrated design in minimizing energy consumption, mitigating disconnectivity, and ensuring consistent communication quality throughout the UAV mission. In particular, the trajectory in the integrated network has a total length of $4,072.94$ meters, involving $1$ coverage hole, whereas the standalone terrestrial network results in a trajectory of $4,354.1$ meters with $5$ coverage holes.

\begin{figure*}[t]
    \centering
    \begin{minipage}{\linewidth}
        \centering
        \includegraphics[width=\linewidth]{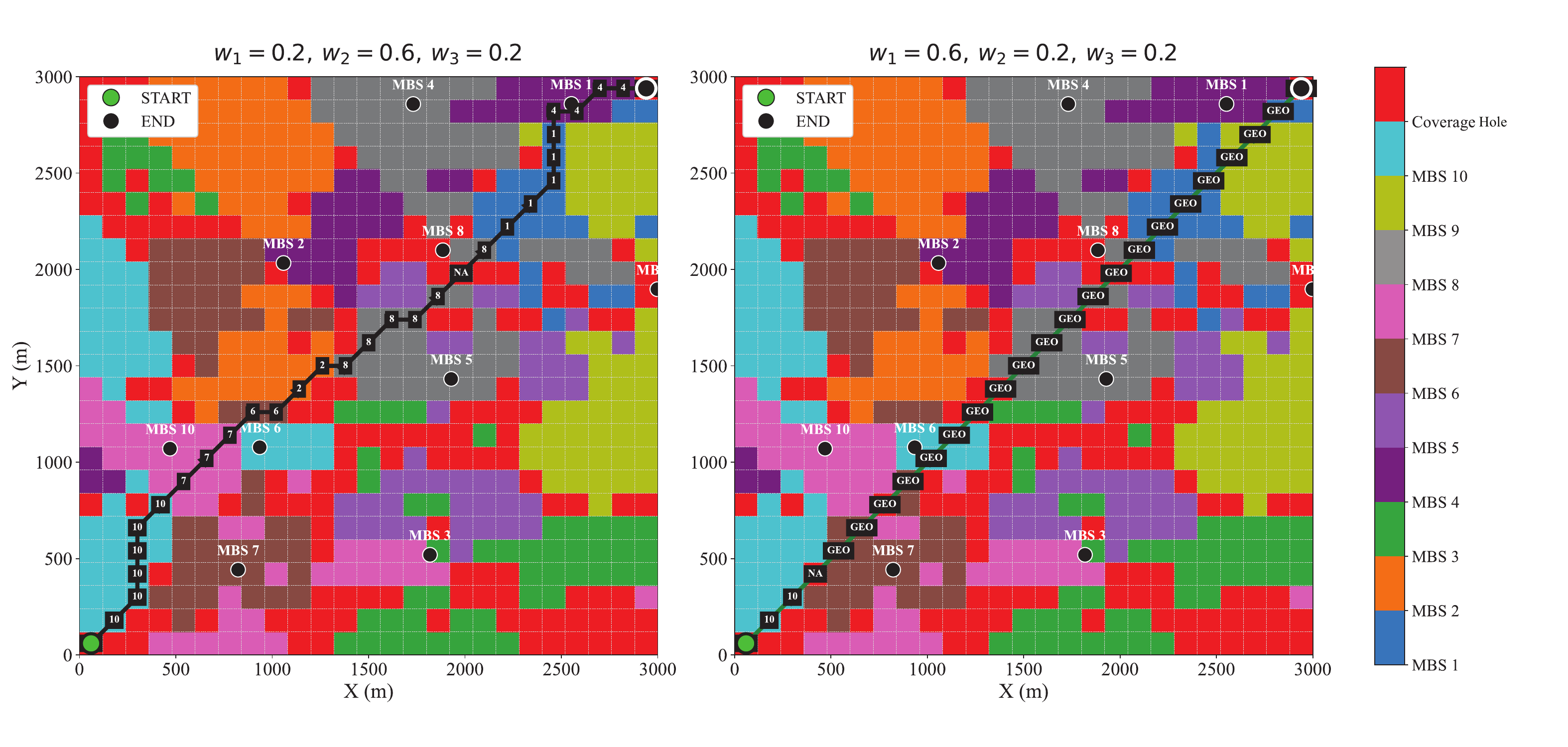}
        \captionof{figure}{UAV trajectory results under two different sets of weight parameters $(w_1, w_2, w_3)$ in the $25 \times 25$ grid environment:\\
        (a) $(0.2, 0.6, 0.2)$, (b) $(0.6, 0.2, 0.2)$.}
        \label{fig:WeightsImpact}
    \end{minipage}
\end{figure*}
Figure.~\ref{fig:WeightsImpact} depicts the UAV trajectory under two different weighting configurations of $(w_1, w_2, w_3)$, namely $(0.2, 0.6, 0.2)$ and $(0.6, 0.2, 0.2)$. When handover cost minimization is prioritized (i.e., $w_2=0.6$), as shown in the left subplot, the UAV follows a path that avoids association with the GEO satellite (due to the higher handover and association cost) and instead remains connected to the terrestrial network with fewer handovers. Consequently, the resulting trajectory has a longer distance, as the UAV sacrifices energy consumption optimality to minimize handover cost. On the other hand, in the right-hand-side plot, where energy consumption is prioritized (i.e., $w_1=0.6$), the UAV follows the shortest path and associates with the GEO satellite to avoid coverage holes. The UAV remains associated with the same link throughout most of the path to further reduce handover cost. In particular, the left-hand-side trajectory experiences a total handover cost of $\sum_{t=1}^T{\eta_{t,(t+1)}} =6.5$, with $1$ coverage hole and a trajectory distance of $4,494.7$ meters, while the right-hand-side trajectory experiences a total handover cost of $10.5$, $1$ coverage hole, and a trajectory distance of $4,072.94$ meters.
% Finally, Fig.~\ref{fig:Scalibility} illustrates the UAV trajectory obtained under three different weight configurations in a $50 \times 50$ grid. This result demonstrates the scalability of the proposed DQN-based algorithm with respect to the grid size $N_G$, confirming its applicability for trajectory and association planning in large-scale regions.
% \begin{figure}[t]
%     \centering
%     \begin{minipage}{\linewidth}
%         \centering
%         \includegraphics[width=0.7\linewidth]{Figures/All_Trajectories_Arrows.eps}
%         \captionof{figure}{\small DQN-based UAV trajectory in a $50 \times 50$ grid.}
%         \label{fig:Scalibility}
%     \end{minipage}
% \end{figure}
\section{Conclusion}\label{Conclusion}
In this work, we investigated the joint UAV trajectory and association planning (JUTAP) problem in an integrated GEO–terrestrial network. To tackle the inherent complexities of this problem, we proposed a DQN–based JUTAP algorithm suitable for large-scale areas with multiple terrestrial MBSs. Simulation results validated the convergence of the proposed algorithm and demonstrated its superior performance in terms of reducing energy consumption, handover cost, and disconnectivity when the GEO satellite complements the terrestrial network in serving the UAV. These results highlight the potential of AI-driven decision-making frameworks for future integrated satellite–terrestrial networks.

\renewcommand\refname{References}

\addcontentsline{toc}{section}{References}
\bibliographystyle{IEEEtran}
\bibliography{references}

@INPROCEEDINGS{mob_uav,
  author={Chowdhury, Md Moin Uddin and Saad, Walid and Güvenç, Ismail},
  booktitle={Proc. IEEE Int. Conf. Commun. Workshops (ICC Workshops)},
  title={{Mobility Management for Cellular-Connected UAVs: A Learning-Based Approach}}, 
  year={2020},
  month={},
  volume={},
  number={},
  pages={1-6},
  doi={10.1109/ICCWorkshops49005.2020.9145089}}

@techreport{ITUR_M2101,
  author       = {{International Telecommunication Union (ITU)}},
  title        = {{Modelling and simulation of {IMT} networks and systems for use in sharing and compatibility studies}},
  institution  = {ITU Radiocommunication Sector (ITU-R)},
  number       = {Recommendation ITU-R M.2101-0},
  month        = feb,
  year         = 2017,
  note         = {02/2017}
}

@INPROCEEDINGS{NesrinArxiv,
  author={Cherif, Nesrine and Nadeem, Qurrat-Ul-Ain},
  booktitle={ICC 2025 - IEEE International Conference on Communications}, 
  title={Merits of Serving {UAVs} via Terrestrial Networks: A Vertical Antenna Radiation Study}, 
  year={2025},
  month={},
  volume={},
  number={},
  pages={1470-1475},
  doi={10.1109/ICC52391.2025.11161080}}

@ARTICLE{Mozaffari,
  author={Mozaffari, Mohammad and Saad, Walid and Bennis, Mehdi and Nam, Young-Han and Debbah, Mérouane},
  journal={IEEE Communications Surveys \& Tutorials}, 
  title={{A Tutorial on {UAVs} for Wireless Networks: Applications, Challenges, and
  Open Problems}}, 
  year={2019},
month={thirdquarter},
  volume={21},
  number={3},
  pages={2334-2360},
  doi={10.1109/COMST.2019.2902862}}

@ARTICLE{cel_uav_aam,
  author={Do, Hieu and Pulgar, Ramon Delgado and Fodor, Gábor and Qi, Zhiqiang},
  journal={IEEE Commun. Stand. Mag.}, 
  title={{Cellular Connectivity for Advanced Air Mobility: Use Cases and Beamforming Approaches}}, 
  year={2024},
  month={Mar.},
  volume={8},
  number={1},
  pages={65-71},
  doi={10.1109/MCOMSTD.0007.2200069}}

@ARTICLE{nesrin,
  author={Cherif, Nesrine and Jaafar, Wael and Yanikomeroglu, Halim and Yongacoglu, Abbas},
  journal={IEEE Trans. Veh. Technol.}, 
  title={{RL-Based Cargo-UAV Trajectory Planning and Cell Association for Minimum Handoffs, Disconnectivity, and Energy Consumption}}, 
  year={2024},
  month={May},
  volume={73},
  number={5},
  pages={7304-7309},
  doi={10.1109/TVT.2023.3340177}}

@ARTICLE{cell_uav,
  author={Zeng, Yong and Lyu, Jiangbin and Zhang, Rui},
  journal={IEEE Wirel. Commun.},
  title={{Cellular-Connected UAV: Potential, Challenges, and Promising Technologies}}, 
  year={2019},
  month={Feb.},
  volume={26},
  number={1},
  pages={120-127},
  doi={10.1109/MWC.2018.1800023}}

@techreport{3gpp_tr36777,
  title        = {{Study on Enhanced LTE Support for Aerial Vehicles (Release 15)}},
  author       = {{3rd Generation Partnership Project; Technical Specification Group Radio Access Network}},
  institution  = {3rd Generation Partnership Project},
  type         = {Technical Report},
  number       = {TR 36.777 V15.0.0},
  year         = {2018},
  month        = {Jan.},
  note         = {Version 15.0.0},
}

@ARTICLE{ai_sat_uav,
  author={Hashima, Sherief and Gendia, Ahmad and Hatano, Kohei and Muta, Osamu and Nada, Mostafa S. and Mohamed, Ehab Mahmoud},
  journal={IEEE Open J. Veh. Technol.},
  title={{Next-Gen UAV-Satellite Communications: AI Innovations and Future Prospects}}, 
  year={2025},
  month={},
  volume={6},
  number={},
  pages={1990-2021},
  doi={10.1109/OJVT.2025.3587028}}

@ARTICLE{energy_aware,
  author={Li, Bowen and Na, Zhenyu and Lin, Bin},
  journal={IEEE Netw.}, 
  title={{UAV Trajectory Planning from a Comprehensive Energy Efficiency Perspective in Harsh Environments}}, 
  year={2022},
  month={Jul./Aug.},
  volume={36},
  number={4},
  pages={62-68},
  doi={10.1109/MNET.006.2100697}}

@INPROCEEDINGS{coverage,
  author={Theile, Mirco and Bayerlein, Harald and Nai, Richard and Gesbert, David and Caccamo, Marco},
  booktitle={Proc. IEEE/RSJ Int. Conf. Intell. Robots Syst. (IROS)},
  title={{UAV Coverage Path Planning under Varying Power Constraints using Deep Reinforcement Learning}},
  year={2020},
  month={},
  volume={},
  number={},
  pages={1444-1449},
  doi={10.1109/IROS45743.2020.9340934}}

@ARTICLE{transfer,
  author={Fontanesi, Gianluca and Zhu, Anding and Arvaneh, Mahnaz and Ahmadi, Hamed},
  journal={IEEE Internet Things J.}, 
  title={{A Transfer Learning Approach for UAV Path Design With Connectivity Outage Constraint}}, 
  year={2023},
  month={Mar.},
  volume={10},
  number={6},
  pages={4998-5012},
  doi={10.1109/JIOT.2022.3220981}}

@INPROCEEDINGS{sat_multi_uav1,
  author={Kourav, Mohit and Singya, Praveen Kumar and Jain, Sandesh},
  booktitle={Proc. IEEE Int. Conf. Interdiscip. Approaches Technol. Manag. Soc. Innov. (IATMSI)}, 
  title={{UAV-Assisted Satellite-Air-Ground Communication: Performance Analysis}}, 
  year={2025},
  month={},
  volume={3},
  number={},
  pages={1-6}}

@ARTICLE{sat_uav2,
  author={Chen, Yu-Jia and Chen, Wei and Ku, Meng-Lin},
  journal={IEEE Commun. Lett.}, 
  title={{Trajectory Design and Link Selection in UAV-Assisted Hybrid Satellite-Terrestrial Network}}, 
  year={2022},
  month={Jul.},
  volume={26},
  number={7},
  pages={1643-1647},
  doi={10.1109/LCOMM.2022.3166961}}

@INPROCEEDINGS{j_allocation,
  author={Hika, Soniya Fufa and Zhu, Likun and Ji, Tianxiang and Du, Haoyu and Chen, Ming and Liu, Jiangtao and Jin, Shihan},
  booktitle={Proc. IEEE 99th Veh. Technol. Conf. (VTC2024-Spring)}, 
  title={{Joint Allocation of Downlink Power, Bandwidth and Timeslots Based on Maximizing Energy Efficiency for GEO Satellite Communication Systems}}, 
  year={2024},
  month={},
  volume={},
  number={},
  pages={1-6},
  doi={10.1109/VTC2024-Spring62846.2024.10683476}}

@ARTICLE{rotor_energy,
  author={Zeng, Yong and Xu, Jie and Zhang, Rui},
  journal={IEEE Trans. Wirel. Commun.}, 
  title={{Energy Minimization for Wireless Communication With Rotary-Wing UAV}}, 
  year={2019},
  month={Apr.},
  volume={18},
  number={4},
  pages={2329-2345},
  doi={10.1109/TWC.2019.2902559}}

\end{document}